\documentclass{elsart5p}
\usepackage{graphics}
\usepackage{graphicx}
\usepackage{amssymb}

\def\12{\frac{1}{2}}

\begin{document}

\begin{frontmatter}


  
  \title{Boson decay and the dynamical structure factor for the $XXZ$ chain at
  finite magnetic field}

\author[AA]{J. Sirker\corauthref{Sirker}},
\ead{j.sirker@fkf.mpg.de}
\author[BB]{R. G. Pereira},
\author[CC]{J.-S. Caux},
\author[CC]{R. Hagemans},
\author[DD]{J. M. Maillet},
\author[EE]{S. R. White},
\author[BB]{I. Affleck}

\address[AA]{Max-Planck-Insitute for Solid State Research, Heisenbergstr. 1,
  70569 Stuttgart, Germany}
\address[BB]{Department of Physics and Astronomy, University of British
  Columbia, Vancouver, British Columbia, Canada V6T 1Z1}
\address[CC]{Institute for Theoretical Physics, University of
  Amsterdam, %Valckenierstraat 65,
  1018 XE Amsterdam, The Netherlands}
\address[DD]{Laboratoire de Physique, \'Ecole Normale Sup\'erieure de Lyon,
  %46 All\'ee d'Italie,
  69364 Lyon CEDEX 07, France}
\address[EE]{Department of Physics and Astronomy, University of California, Irvine
  CA 92697, USA}

\corauth[Sirker]{Corresponding author. Tel: (+49) 711689-1649 fax: (+49)
711689-1702}

\begin{abstract}
  We study the longitudinal dynamical structure factor $S^{zz}(q,\omega)$ for
  the anisotropic spin-$1/2$ ($XXZ$) chain at finite magnetic field using
  bosonization. The leading irrelevant operators in the effective bosonic
  model stemming from band curvature describe boson decay processes and lead
  to a high-frequency tail and a finite width $\gamma_q$ of the on-shell peak
  for $S^{zz}(q,\omega)$.  We use the Bethe ansatz to show that $\gamma_q\sim
  q^2$ for $q\ll 1$ and to calculate the amplitudes of the leading irrelevant
  operators in the effective field theory.
\end{abstract}


\end{frontmatter}

% main text

\section{Introduction}
%\label{}
The Hamiltonian of the $XXZ$ spin chain is given by
\begin{equation}
H=J\sum_{j=1}^{N}\left[S_{j}^{x}S_{j+1}^{x}+S_{j}^{y}S_{j+1}^{y}+\Delta
  S_{j}^{z}S_{j+1}^{z}-hS_{j}^{z}\right]\; .
\label{XXZ}
\end{equation}
Here $J$ is the exchange coupling, $\Delta$ the anisotropy, $h$ a magnetic
field and $N$ the number of sites. The isotropic model ($\Delta = 1$) has
direct applications to compounds like Sr$_2$CuO$_3$ where copper-oxygen chains
form along one crystal axis and the coupling between these chains is very weak
making them effectively one dimensional. Furthermore, the system (\ref{XXZ})
can be mapped exactly onto interacting spinless fermions using the
Jordan-Wigner transformation. In this language $\Delta$ parametrizes a
nearest-neighbor density-density interaction and $h$ becomes the chemical
potential. The fermionic version of (\ref{XXZ}) is often used as a simple
model for quantum wires \cite{Glazman}.  

The $XXZ$ model is exactly solvable by means of Bethe Ansatz (BA). However,
the calculation of correlation function by BA is a difficult task and only
limited results have been obtained so far. On the other hand, it is known that
the low-energy effective theory for the Hamiltonian (\ref{XXZ}) is just a free
boson model. This so called Luttinger liquid Hamiltonian $H_{LL}$ is obtained
by linearizing the dispersion around the two Fermi points, going to the continuum
limit and reexpressing the right- and left-moving fermions by bosons. Ignoring
irrelevant operators this leads to 
\begin{equation}
%% H_{LL} = v\sum_{q>0} q\l[a_q^{R\dagger} a_q^R + a_q^{L\dagger} a_q^L \r] \;
%% ,
H_{LL} = \frac{v}{2}\int dx \left[\Pi^2+\left(\partial_x\phi\right)^2\right] \, .
\label{HLL}
\end{equation}
Here, $\phi(x)$ is a bosonic field and $\Pi(x)$ its conjugated momentum
satisfying $[\phi(x),\Pi(y)]=i\delta(x-y)$. As a consequence of the linearized
dispersion the Hamiltonian (\ref{HLL}) is Lorentz invariant and a boson with momentum
$|q|$ always carries energy $\omega = v|q|$. In the following we will
investigate corrections to (\ref{HLL}) due to band curvature and consequences
for the lineshape of the dynamical structure factor.
\section{The dynamical structure factor}
The longitudinal dynamical structure factor is defined by
\begin{eqnarray}
S^{zz}\left(q,\omega\right)&=&\frac{1}{N}\sum_{j,j^{\prime}=1}^{N}e^{-iq\left(j-j^{\prime}\right)}\int_{-\infty}^{+\infty}
dt\,
e^{i\omega t}\left\langle
  S_{j}^{z}\left(t\right)S_{j^{\prime}}^{z}\left(0\right)\right\rangle
\nonumber \\
&=& \frac{2\pi}{N}\sum_{\alpha}\left|\left\langle
    0\left|S_{q}^{z}\right|\alpha\right\rangle
\right|^{2}\delta\left(\omega-E_{\alpha}\right) \; .
\label{strucFac}
\end{eqnarray}
Here $S_{q}^{z}=\sum_{j}S_{j}^{z}e^{-iqj}$ and $\left|\alpha\right\rangle $ is
an eigenstate with energy $E_{\alpha}$ above the ground state energy.  For a
finite system, $S^{zz}\left(q,\omega\right)$ at fixed $q$ is a sum of
$\delta$-peaks at the energies of the eigenstates.  In the thermodynamic limit
$N\rightarrow\infty$, the spectrum is continuous and
$S^{zz}\left(q,\omega\right)$ becomes a smooth function of $\omega$ and $q$.
Due to Lorentz invariance
$S^{zz}\left(q,\omega\right)=K\left|q\right|\delta\left(\omega-v\left|q\right|\right)$
for the Luttinger model (\ref{HLL}). 

\begin{figure}[!ht]
\begin{center}
\includegraphics*[width=0.99\columnwidth]{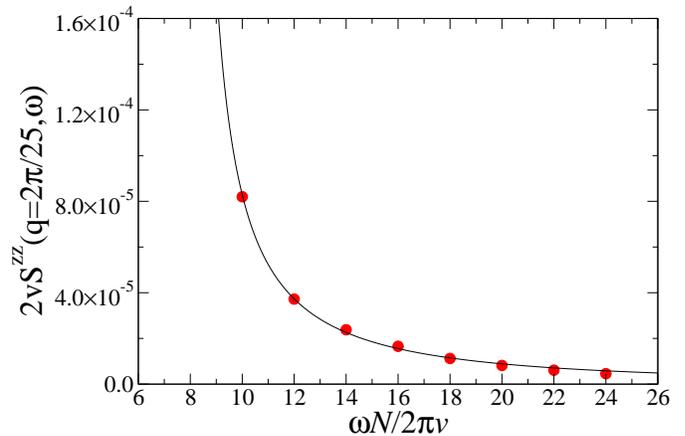}
\end{center}
\caption{Analytical result for the high-frequency tail (\ref{fF_tail}) (black
  solid line) compared to a BA calculation for a chain with $N=200$
  (red dots).  Here $\Delta = 0.25$ and the magnetization $\langle s^z\rangle=-0.1$.}
\label{fig1}
\end{figure}

For $h\neq 0$ particle-hole symmetry is broken and the leading irrelevant
operators allowed by symmetry are the dimension-three operators $\sim
(\partial_x\phi_{R,L})^3$. More precisely, we consider the following
correction to (\ref{HLL})
\begin{eqnarray}
 \label{correct}
 \delta H &=& \int dx \,\bigg\{
   \eta_{-}\left[\left(\partial_{x}\phi_{L}\right)^{3}-\left(\partial_{x}\phi_{R}\right)^{3}\right] \\
&+& \eta_{+}\left[\left(\partial_{x}\phi_{L}\right)^{2}\partial_{x}\phi_{R}-\left(\partial_{x}\phi_{R}\right)^{2}\partial_{x}\phi_{L}\right]\bigg\}
   \nonumber \; ,
\end{eqnarray}
 where $\phi_{R,L}$ are the right- and left-moving components of the bosonic
 field with $\phi=(\phi_L-\phi_R)/\sqrt{2}$. This correction is the bosonized
 form of the leading band curvature term $\omega(q) = \pm vq+ q^2/(2m)+\cdots$.
 Note, that (\ref{correct}) corresponds to interaction vertices allowing a boson
 to decay into two bosons. The $\eta_{+}$-interaction allows for intermediate
 states with one right- and one left-moving boson, which together can carry
 small momentum but high energy $\omega\gg v\left|q\right|$ thus giving rise
 to a high-energy tail in $S^{zz}(q,\omega)$. The $\eta_{-}$-interaction, on
 the other hand, will influence the on-shell part $\omega\sim vq$. If
 $\gamma_q \ll \omega -v|q|\ll J$, where $\gamma_q$ is the width if the
 on-shell peak, then the terms in (\ref{correct}) can be treated in perturbation
 theory and we find a high-frequency tail \cite{SirkerPereira}
\begin{equation}
\delta
S^{zz}_{\eta_+}\left(q,\omega\right)=\frac{K\eta_{+}^{2}q^{4}}{v\pi}\,\frac{\theta\left(\omega-v\left|q\right|\right)}{\omega^{2}-v^{2}q^{2}}.
\label{fF_tail}
\end{equation}
The parameters $\eta_\pm$ can be related to the change in the velocity $v$ and
the Luttinger parameter $K$ when varying $h$ and we find $J\eta_-(h_0) =
\sqrt{2\pi/K}v^2(a+b/2)/6$ and $J\eta_+(h_0) = \sqrt{2\pi/K}v^2b/4$ where
$a=v^{-1}\partial v/\partial h |_{h=h_0}$ and $b=K^{-1}\partial K/\partial h
|_{h=h_0}$ \cite{SirkerPereira}. Due to the integrability of the model it is
possible to obtain $v(h)$, $K(h)$ for all anisotropies and fields by BA so
that (\ref{fF_tail}) does not contain any free parameters. In Fig.~\ref{fig1}
this analytical result is compared to a BA calculation for a chain with
$N=200$ sites.

\begin{figure}[!ht]
\begin{center}
\includegraphics*[width=0.99\columnwidth]{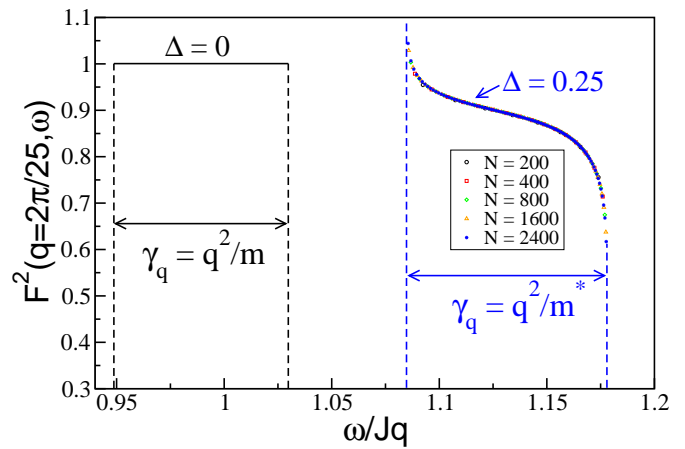}
\end{center}
\caption{Exact solution for $F(q,\omega)$ at the free fermion point $\Delta =
  0$ compared to numerical BA data at $\Delta = 0.25$ for different finite
  size systems with $\langle s^z\rangle=-0.1$ in both cases. The effective
  mass in the free fermion case is given by $m=(J\cos k_F)^{-1}$ where $k_F$
  is the Fermi momentum. For small finite $\Delta$ the effective mass can be
  obtained by BA and is given by $m^* = m/(1+2\Delta\sin k_F/\pi)$
  \cite{SirkerPereira}.}
\label{fig2}
\end{figure}

Perturbation theory in $\eta_-$, on the other hand, produces terms which become
increasingly singular at $\omega\sim vq$. Therefore the whole series has to be
summed up to produce a finite result \cite{PereiraSirkerPrep}. Here we only
present BA results for the form factors $F(q,\omega)\equiv \langle
0|S^z_q|\alpha\rangle$ (see (\ref{strucFac})) for $\omega\sim vq$ in
Fig.~\ref{fig2}.
Note that the width $\gamma_q$ of the on-shell peak for finite $\Delta$ is
$\sim q^2$ as in the free fermion case ($\Delta = 0$) but the effective mass
is renormalized. For finite $\Delta \ll 1$, Pustilnik {\it et al.}
\cite{Glazmannew} have shown that $S^{zz}(q,\omega)$ exhibits power laws at
the upper and lower threshold of the on-shell peak with exponents depending on
$\Delta$ {\it and} $q$. Their results are consistent with our numerical data
shown in Fig.~\ref{fig2}.
%% \section{Summary}

\section{Acknowledgement}
This research was supported by CNPq through Grant
No.~200612/2004-2 (R.G.P), the DFG (J.S.), FOM
%% Stichting voor Fundamenteel Onderzoek der Materie 
(J.-S.C.), CNRS and the EUCLID network (J.M.M.), the NSF under DMR 0311843
(S.R.W.), and NSERC (J.S., I.A.) and the CIAR (I.A.).

%% \bibliography{Literatur}

\begin{thebibliography}{99}

\bibitem{Glazman} M. Pustilnik, E. G. Mishenko, L. I. Glazman, and
  A. V. Andreev, Phys. Rev. Lett. {\bf 91} (2003) 126805.
\bibitem{SirkerPereira} R. G. Pereira, J. Sirker, J.-S. Caux, R. Hagemans,
  J. M. Maillet, S. R. White, and I. Affleck, Phys. Rev. Lett. {\bf 96} (2006)
  257202.
\bibitem{PereiraSirkerPrep} R. G. Pereira, J. Sirker, J.-S. Caux, R. Hagemans,
  J. M. Maillet, S. R. White, and I. Affleck, in preparation.
\bibitem{Glazmannew} M. Pustilnik, M. Khodas, A. Kamenev, and L. I. Glazman,
  Phys. Rev. Lett.  {\bf 96} (2006) 196405.



 \end{thebibliography}

\end{document}